\documentclass[twocolumn,showpacs,superscriptaddress,amsmath,amssymb]{revtex4}

\usepackage{graphicx}
\usepackage{dcolumn}
\usepackage{bm}

\begin{document}

\title{
Unusual pseudogap-like features observed in iron oxypnictide 
superconductors}

\author{Y.~Ishida}
\affiliation{RIKEN SPring-8 Center, Sayo, Sayo, Hyogo 679-5148, 
Japan}

\author{T.~Shimojima}
\affiliation{ISSP, University of Tokyo, Kashiwa-no-ha, Kashiwa, 
Chiba 277-8561, Japan}

\author{K.~Ishizaka}
\affiliation{ISSP, University of Tokyo, Kashiwa-no-ha, Kashiwa, 
Chiba 277-8561, Japan}
\affiliation{CREST, Japan Science and Technology Agency, 
Tokyo 102-0075, Japan}

\author{T.~Kiss}
\author{M.~Okawa}
\affiliation{ISSP, University of Tokyo, Kashiwa-no-ha, Kashiwa, 
Chiba 277-8561, Japan}

\author{T.~Togashi}
\affiliation{RIKEN SPring-8 Center, Sayo, Sayo, Hyogo 679-5148, 
Japan}

\author{S.~Watanabe}
\affiliation{ISSP, University of Tokyo, Kashiwa-no-ha, Kashiwa, 
Chiba 277-8561, Japan}

\author{X.-Y.~Wang}
\author{C.-T.~Chen}
\affiliation{Technical Institute of Physics and Chemistry, 
Chinese Academy of Science, Zhongguancun, Beijing 100080, China}

\author{Y.~Kamihara}
\affiliation{ERATO-SORST, JST, Frontier Research Center, Tokyo Institute of 
Technology, Mail Box S2-13, 4259 Nagatsuta, Midori-ku, Yokohama 226-8503, 
Japan}
\affiliation{Frontier Research Center, Tokyo Institute of 
Technology, Mail Box R3-1, 4259 Nagatsuta, Midori-ku, Yokohama 226-8503, 
Japan}

\author{M.~Hirano}
\affiliation{ERATO-SORST, JST, Frontier Research Center, Tokyo Institute of 
Technology, Mail Box S2-13, 4259 Nagatsuta, Midori-ku, Yokohama 226-8503, 
Japan}
\affiliation{Materials and Structures Laboratory, Tokyo Institute of 
Technology, Mail Box R3-1, 4259 Nagatsuta, Midori-ku, Yokohama 226-8503, 
Japan}

\author{H.~Hosono}
\affiliation{ERATO-SORST, JST, Frontier Research Center, Tokyo Institute of 
Technology, Mail Box S2-13, 4259 Nagatsuta, Midori-ku, Yokohama 226-8503, 
Japan}
\affiliation{Frontier Research Center, Tokyo Institute of 
Technology, Mail Box R3-1, 4259 Nagatsuta, Midori-ku, Yokohama 226-8503, 
Japan}
\affiliation{Materials and Structures Laboratory, Tokyo Institute of 
Technology, Mail Box R3-1, 4259 Nagatsuta, Midori-ku, Yokohama 226-8503, 
Japan}

\author{S.~Shin}
\affiliation{ISSP, University of Tokyo, Kashiwa-no-ha, Kashiwa, 
Chiba 277-8561, Japan}
\affiliation{RIKEN SPring-8 Center, Sayo, Sayo, Hyogo 679-5148, 
Japan}
\affiliation{CREST, Japan Science and Technology Agency, 
Tokyo 102-0075, Japan}

\date{\today}

\begin{abstract}
We have performed a temperature-dependent angle-integrated 
laser photoemission study of iron oxypnictide superconductors 
LaFeAsO:F and LaFePO:F exhibiting critical transition temperatures 
($T_c$'s) of 26 K and 5 K, respectively. 
We find that high-$T_c$ LaFeAsO:F exhibits a temperature-dependent 
pseudogap-like feature extending over $\sim$0.1 eV about 
the Fermi level at 250 K, 
whereas such a feature is absent in low-$T_c$ LaFePO:F. 
We also find 
$\sim$20-meV pseudogap-like features and signatures of superconducting gaps 
both in LaFeAsO:F and LaFePO:F. 
We discuss the possible origins of the 
unusual pseudogap-like features through comparison with the high-$T_c$ 
cuprates. 
\end{abstract}



\maketitle

Realization of room-temperature superconductivity is one of the 
ultimate goals in the field of materials science. 
In order to achieve a high critical-transition temperature ($T_c$), 
one needs to elucidate how materials can overcome 
$T_c\sim$ 40 K, a so-called 
BCS limit predicted from a theory of phonon-meditated superconductivity 
\cite{McMillan}. 
The highest $T_c$ 
reported 
to date is $\sim$160 K in the cuprates 
\cite{150K, 160K}, and their high-$T_c$ mechanism is 
believed to be linked to mysterious pseudogaps 
observed in the abnormal metallic phase at 
$T>T_c$ \cite{Timusk}. 
Recently, transition-metal oxypnictides 
composed of alternate stacking of 
\textit{Ln}$_2$O$_2$ layers and 
\textit{T}$_2$\textit{Pn}$_2$ layers 
(\textit{Ln}: lanthanide; 
\textit{T}: Fe or Ni; \textit{Pn}: P or As) 
were identified as new high-$T_{\rm c}$ materials 
\cite{LOFFA_Kamihara, LOFFP_Kamihara, LOFFP_Liang, LONiP_Watanabe}
that exceed the BCS limit up to 55 K
\cite{Pressure_Nature, XHChen, Sm_55K, Ce_41K}. 
The mechanism of the high $T_c$'s has been the 
subject of strong debate: it is unclear 
whether the mechanism is a parallel case to the high-$T_c$ cuprates 
or a completely new case. 

Photoemission spectroscopy (PES) is a powerful tool to investigate 
the electronic structures of materials. Angle-integrated 
PES studies on FeAs-based superconductors have revealed 
pseudogap features with energies ranging from 15 to 40 meV 
at $T\lesssim$ 100 K \cite{Sato_JPSJ, Liu, Feng, Aiura, Garcia}. 
However, subsequent angle-resolved PES studies 
on iron oxypnictides \cite{Shen, Ding} 
showed that the Fermi surfaces in the normal state 
were fully preserved, contrasted to the 
doping-dependent pseudogaps of the cuprates that 
cause the Fermi-surface sectors to vanish \cite{WSLee}. 
Herein, we present a 
comparative electronic-structure 
study of optimally-F-doped LaFeAsO:F 
(As26) \cite{LOFFA_Kamihara} and LaFePO:F (P05) 
\cite{LOFFP_Kamihara} 
exhibiting $T_c$'s of 26 and 5 K, respectively. 
Through investigating the $T$ dependence of the 
angle-integrated spectra recorded by a laser PES system \cite{Kiss2008}, 
we find that even the low-$T_c$ P05 
exhibits a pseudogap feature similar to those of 
the high-$T_c$ FeAs system 
\cite{Sato_JPSJ, Liu, Feng, Aiura, Garcia}. We also find 
a large $T$-dependent pseudogap feature specific to 
high-$T_c$ As26 manifesting at high $T$'s. 
Our study shows that pseudogaps do exist in the 
iron oxypnictides, but they are qualitatively different from the 
doping-dependent pseudogaps of the cuprates. 

Polycrystalline As26 and P05 
were fabricated as described elsewhere 
\cite{LOFFA_Kamihara, LOFFP_Kamihara, LOFFA_Chara}. 
The F contents substituting for O in As26 and in P05 were 
estimated to be 11 and 6 \%, respectively. 
Laser PES 
measurements with an excitation energy of $h\nu$ = 6.994 eV 
were performed 
at ISSP, University of Tokyo \cite{Kiss2008}. 
The base pressure of the spectrometer was 
$<$2$\times$10$^{-11}$ Torr throughout the measurements. 
The Fermi level ($E_{F}$) 
was carefully calibrated within an accuracy of 
$\pm$$\textless$0.01 meV by recording Fermi cutoffs of 
Ag or Au in electrical 
contact with the sample and the analyzer. 
The spectra were recorded in an 
angle-integrated mode and typical energy resolutions 
during As26, P05 and 
Au measurements were 3, 3, and 1 meV, respectively. 

\begin{figure}[htb]
\begin{center}
\includegraphics[width=8.7 cm]{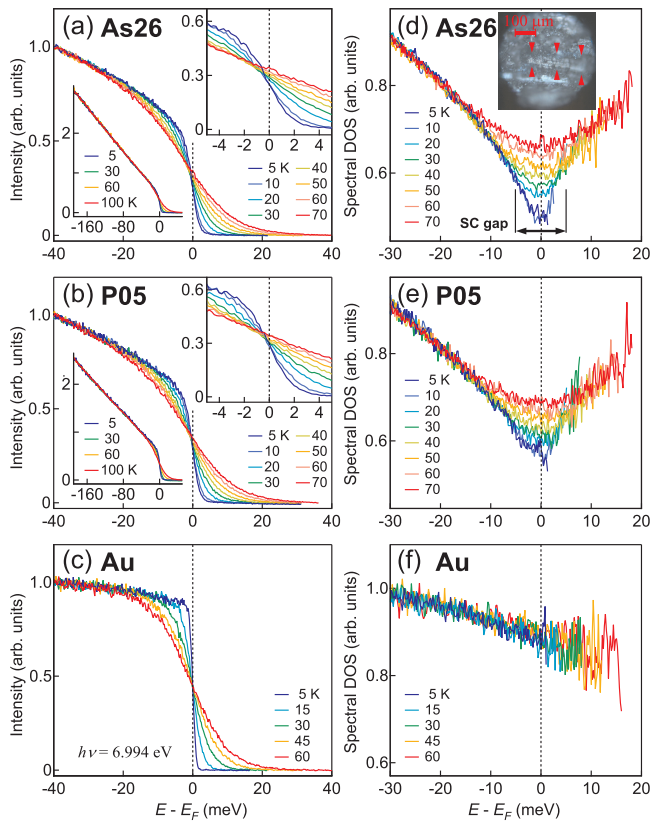}
\caption{\label{fig1} 
Temperature dependence of the laser PES spectra of As26 (a), P05 (b) 
and Au (c). The spectra were normalized to the intensity at 40 meV 
below $E_F$. 
The upper-right insets in (a) and (b) show enlarged plots near $E_F$. 
The lower-left insets in (a) and (b) show wide-range spectra 
recorded at $T\le$ 100 K and 
normalized to the intensity at -40 meV. 
The spectral DOSs of As26, P05 and Au derived from the spectra shown in 
the main panels of (a-c) are shown in (d-f), respectively. 
The inset in (c) shows a plate-like area of As26 
(indicated by arrows) on which we recorded the spectra. 
}
\end{center}
\end{figure}

We adopted ``soft scraping" for the sample surface preparation, 
as we could obtain signatures of superconducting gaps 
in the spectra (described later) through this method. 
First, we polished the sample and made the surface convex. 
Then, the sample was introduced into the PES spectrometer 
and was softly scraped once with a diamond file at 150 K. 
Since the diamond file had a point contact with the convex surface of the 
sample, soft scraping resulted in a couple of lines of scratches 
on the surface. With the aid of a 
large-magnification CCD camera allowing us to view the 
laser spot 
position on the surface (spot diameter was $\lesssim$300 $\mu$m) 
\cite{Kiss2008}, 
we searched along the scratch for 
the highest photoemission count rate. After the 
measurements, we checked the surface by 
an optical microscope and found a plate-like area of 
$\sim$200 $\times$ 50 $\mu$m$^2$ [image inset in Fig.\ \ref{fig1}] 
at the exact position where we had recorded the spectra. 
We note that the spectra recorded on fractured surfaces mostly showed 
$T$ dependence of a Fermi-Dirac function even below $T_c$, 
an unusual situation for a superconductor. 
We speculate that fracturing mostly resulted in 
exposure of metallic grain boundaries in the present sample. 

The main panels of Fig.\ \ref{fig1}(a) and (b) show 
laser PES spectra of As26 
and P05, respectively, recorded at $T\le$ 70 K. 
Here, the spectra were normalized to the intensity at 
40 meV below $E_{F}$, since the spectra showed 
little change at $ E-E_{F}\textless$ -40 meV 
at $T\leq$ 100 K as shown in the lower-left insets 
in Fig.\ \ref{fig1}(a) and (b). 
The spectra of both As26 and P05 show almost linear 
slopes towards $E_{F}$, consistent with the calculations 
based on local-density approximation \cite{Lebegue, Singh}. 
In the magnified plots of the near-$E_F$ spectra of As26 and P05 
shown in the upper-right insets in Fig.\ \ref{fig1}(a) and (b), 
respectively, one can see that the intensity at $E_{F}$ 
gradually becomes weak with decreasing $T$. 
The intensity variation at $E_{F}$ is observed already 
at between 70 and 60 K, considerably higher than 
the $T_{\rm c}$. 
This is in strong contrast to the $T$ dependence of 
a ``normal" metal as demonstratively shown by that of Au in 
Fig.\ \ref{fig1}(c).
Clearly, all the spectra of Au intersect at a single point 
at $E_{F}$ since 
their $T$ dependence obeys that of a 
Fermi-Dirac distribution function 
which takes a $T$-independent value of 1/2 at $E_{F}$. 
Thus, the metallic phase above the superconducting phase of 
As26 and P05 is pseudogapped, that is, 
the near-$E_{F}$ spectral weight is mysteriously depressed 
with decreasing $T$ 
even in the absence of an apparent order parameter.

In order to investigate the energy scale of the pseudogaps at $T\le$ 70 K, 
we have divided the laser PES spectra of As26 and P05 by the 
Fermi-Dirac distribution function 
convoluted with a Gaussian corresponding to the instrumental resolution. 
Through this procedure, one can isolate the $T$ dependence of 
spectral density of states (DOS) from that of the 
Fermi-Dirac distribution function. 
The spectral DOSs of As26 and P05 thus obtained are 
shown in Fig.\ \ref{fig1}(d) and (e), respectively. 
One can see that the spectral DOS of As26 
is depressed with decreasing $T$ below 70 K 
in the energy range of $\pm$$\sim$20 meV, and we identify it 
to the 15\,-\,40-meV pseudogaps 
reported in the angle-integrated PES studies on the FeAs system 
\cite{Sato_JPSJ, Liu, Feng, Aiura, Garcia}. 
Surprisingly, a $T$ dependence similar to 
the $\sim$20-meV pseudogap of As26 is present 
even in the spectral DOS of low-$T_c$ P05 [Fig.\ \ref{fig1}(e)]. 
This indicates that the $\sim$20-meV pseudogaps 
at $T\lesssim$100 K are commonly observed in 
the iron oxypnictides and are irrelevant to $T_c$. 
The anomalous pseudogap features of As26 and P05 
are in clear contrast to the $T$-independent spectral DOS of 
Au shown in Fig.\ \ref{fig1}(f) 
representing a ``normal" metal. 

\begin{figure}[htb]
\begin{center}
\includegraphics[width=8.7 cm]{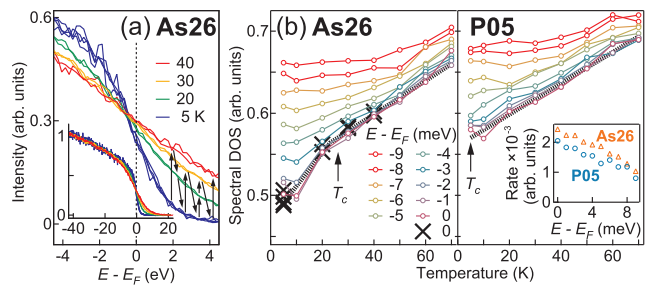}
\caption{\label{fig2} 
Superconducting-gap feature of As26. (a) 
Near-$E_F$ spectra of As26 recorded 
during a $T$ cycling. 
The sequence of the $T$ cycling is indicated by arrows. 
The inset shows the spectra in a wide energy range. (b) 
Spectral DOS weights of 
As26 (left panel) and P05 (right panel) 
at several energies, derived from the 
data shown in Fig.\ \ref{fig2} (circles) 
and in Fig.\ \ref{fig3}(a) (crosses). 
The dotted lines are guide to the eye. 
The inset shows the depression rate of the 
spectral DOS weights with decreasing $T$ from 70 to 30 K.  
}
\end{center}
\end{figure}

\begin{figure}[htb]
\begin{center}
\includegraphics[width=8.7 cm]{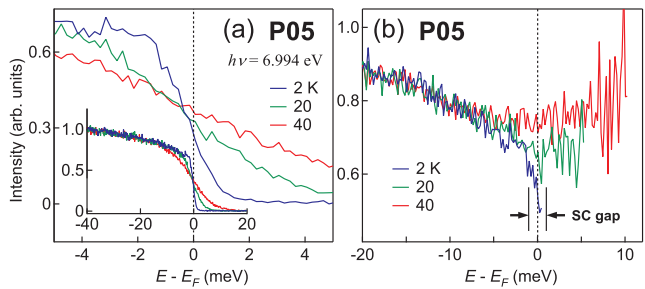}
\caption{\label{fig3} 
Superconducting-gap feature of P05. 
(a) Laser PES spectra recorded on fractured P05. 
The inset shows wide-energy-range spectra normalized to the 
intensity at 40 meV below $E_{F}$. 
(b) Spectral DOS of P05 showing signatures of the 
superconducting gap and the pseudogap. 
}
\end{center}
\end{figure}

Comparing the spectral DOSs of As26 and P05 at 
5 $\leq T\leq$ 20 K shown in Fig.\ \ref{fig1}(d) and (e), respectively, 
one notices that the spectral weight at $\sim$5 meV about $E_{F}$ of As26 
is more depressed than that of P05. 
The rapid depression of the 
spectral DOS weight of As26 below $T_c$ was reproductively observed in 
a $T$-cycling measurement conducted separately 
[Fig.\ \ref{fig2}(a)], 
and we identify it to the opening of a superconducting gap. 
The line shapes of the spectral DOSs of As26 at $T$ = 5 and 10 K 
are almost identical [Fig.\ \ref{fig1}(d)], 
indicating that the superconducting gap is mostly opened 
at $T$ = 10 K, reasonable for a $T_{\rm c}$ = 26 K sample. 
The signature of the superconducting gap of As26 is further highlighted 
in Fig.\ \ref{fig2}(b): 
the spectral DOS weights of both As26 and P05 
in the normal state are $T$-linear due to 
the $\sim$20 meV pseudogap, 
but those of As26 for 
$E-E_{F}\gtrsim$ -5 meV 
exhibit rapid depressions below $T_c$ =  26 K. 
The spectra of P05 in the superconducting state 
were recorded by another laser PES spectrometer 
designed for low-$T$ measurements 
as shown in Fig.\ \ref{fig3}(a) \cite{lowT_P05}. 
The spectral DOS of P05 at 2 K shown in Fig.\ \ref{fig3}(b) 
steepens towards $E_{F}$ at $E-E_{F}\gtrsim$-1 meV, 
attributable to the superconducting gap of P05 [please 
compare it with the $T$-independent spectral DOS of Au shown in Fig.\ 
\ref{fig1}(f) serving as a control of the laser-PES setup]. 
The signature of the superconducting gap 
occuring at $\sim$1 meV about $E_{F}$ is reasonably smaller 
than that of As26 concerning the $T_c$. 
However, the spectra recorded in the superconducting states of 
As26 and P05 fail 
to exhibit superconducting peaks and are not fully gapped. 
The finite intensity at $E_F$ in the superconducting state may be caused by 
a non-superconducting volume \cite{Evtushinsky}, and threfore, 
this issue remains to be clarified in future.

\begin{figure}[htb]
\begin{center}
\includegraphics[width=8.7 cm]{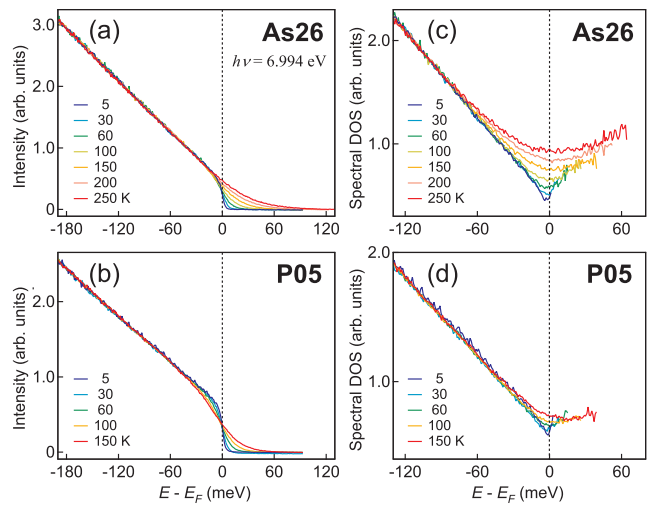}
\caption{\label{fig4} 
Laser PES spectra of As26 (a) and P05 (b) recorded in a 
wide-$T$ range, and the spectral DOSs of As26 (c) and P05 (d) derived 
from the spectra shown in (a) and (b), respectively. 
The spectra were normalized to the 
intensity at $E$\,-\,$E_F$ = -180 meV, well beyond the energy region 
affected by the pseudogaps and the Fermi-Dirac broadenings. }
\end{center}
\end{figure}

Although the pseudogap features of high-$T_c$ As26 
and low-$T_c$ P05 were nearly 
identical at $T\le$ 70 K, 
prominent difference between the two emerges at high $T$'s. 
Figure \ref{fig4}(a) and (b) show spectra of As26 and P05, 
respectively, recorded in a wide-$T$ range, and 
Fig.\ \ref{fig4}(c) and (d) show spectral DOSs of As26 and P05, 
respectively. There is a large 
pseudogap feature in the spectral DOS of As26 
that becomes as large as 
$\sim$0.1 eV at 250 K but that shrinks with decreasing $T$. 
On the other hand, the pseudogap feature of P05 is 
confined within $\pm$$\sim$20 meV about $E_F$ at $T\le$ 150 K. 
We also performed He I PES measurements 
and confirmed that 
the large $T$-dependent pseudogap was present in As26 but not in P05, 
even though the spot diameter of the He lamp 
was as large as $\sim$5 mm \cite{Ishida_Suppl}. 
Ou {\it et al}.\ also observed a large $T$-dependent pseudogap feature 
in SmFeAsO:F, although they could not exclude extrinsic 
polycrystalline effects for this feature \cite{Feng}. 
Since the signature of the 
superconducting gap in 
As26 [Fig.\ \ref{fig2}] 
serves as a credit that the spectra are reflecting the 
bulk electronic structure, 
we believe that the large $T$-dependent pseudogap 
is an intrinsic feature of As26.

Since the depression rates of the spectral-DOS weights 
with decreasing $T$ from 70 to 30 K in As26 and P05 
are nearly identical [inset in Fig.\ \ref{fig2}(b)], we consider 
that the $\sim$20-meV pseudogap feature observed in As26 and P05 
at $T\le$ 70 K have a common origin. 
The $\sim$20-meV pseudogap is presumably 
unrelated to magnetism concerning the diversity of 
magnetic suceptibilities between As26 and P05
(the former shows a small paramagnetic behavior 
\cite{LOFFP_Kamihara} 
whereas the latter shows Curie-Weiss-like 
paramagnetism \cite{LOFFA_Kamihara}). 
Since $\sim$20 meV coincides with the energy scale of 
local Fe-As lattice vibrations 
which are 
softened compared to what is expected from 
local-density approximation calculations \cite{PhononDOS, Baron}, 
the $\sim$20-meV pseudogaps could be attributed to electronic excitations 
coupled to such vibronic modes. 
In fact, Chainani {\it et al}.\ \cite{Chainani} attributed 
the $\sim$70-meV pseudogap of a phonon-meditated superconductor, 
Ba$_{1-x}$K$_x$BiO$_3$ \cite{Chainani}, 
to strong electron-phonon coupling, since $\sim$70 meV is the 
energy of a breathing phonon mode. 
Even though isotropic electron-phonon couplings have been predicted to be 
weak in the iron oxypnictides \cite{Boeri}, 
strong anisotropic 
couplings still have a chance to exist \cite{Eschrig}. 
It is also interesting to note that Hashimoto {\it et al}.\ 
\cite{Hashimoto} recently 
introduced a $\sim$70-meV pseudogap 
that occurs in the angle-integrated PES spectra of hole-doped 
La$_2$CuO$_4$ in the entire hole concentrations \cite{Hashimoto}. 
This new $\sim$70-meV pseudogap 
superimposed on the doping-dependent pseudogap of the cuprates 
\cite{Hashimoto} 
may share a common point to 
the $\sim$20-meV pseudogaps presented herein, 
since both are independent of $T_c$ or of 
doping. Hashimoto {\it et al}.\ proposed that 
the $\sim$70-meV pseudogap could be linked to 
$T$ dependence of $\sim$70-meV kinks 
ubiquitously observed 
in the nodal quasiparticle dispersions in the cuprates 
\cite{Lanzara}. 
In fact, an angle-resolved PES study on 
(Sr,Ba)$_{1-x}$(K,Na)$_x$Fe$_2$As$_2$ reported dispersion 
kinks at 15\,-\,50 meV below $E_F$ \cite{Hasan}, 
corroborating with the energy scale of the $\sim$20-meV pseudogap. 

On the other hand, the large $T$-dependent pseudogap of high-$T_c$ As26, 
which is considered to be superimposed on the $\sim$20-meV pseudogap, 
cannot be attributed to precursors of the superconducing gap, 
since the energy scale of $\sim$0.1 eV at 250 K is 
exceedingly larger than the superconducting gap size of $\sim$5 meV. 
One scenario for the large $T$-dependent pseudogap of As26 is that the 
electronic  excitations are coupled to spin fluctuations in the 
normal state of As26, since the magnetic correlations 
of the antiferromagnetic parent compound \cite{SDW_Neutron, Inelastic} 
may persist in the F-doped superconductor. 
However, it is difficult to understand the $T$ dependence 
of the pseudogap in the spin-fluctuation scenario 
unless we invoke a situation such as the coupled-mode energy of the spin 
fluctuation becomes large at high $T$'s. 
Alternatively, we recall that similar $T$-dependent pseudogaps 
have been observed in Kondo insulators \cite{Kumigashira, Susaki} and 
in thermoelectric skutterudites showing little 
signatures of electron correlations \cite{Ishii}. 
Although the origin of the $T$-dependent pseudogaps in these materials 
is also unclear \cite{Kumigashira, Susaki, Ishii}, the common place 
is that they are narrow-gap \cite{Kumigashira, Susaki} 
or semimetallic \cite{Ishii} materials 
having low carrier concentrations. 
Therefore, the $T$-dependent pseudogap of As26 may be a fingerprint that 
As26 is also a low-carrier density semimetal \cite{Ishida_Suppl}, 
consistent with a picture given in ref.\ \cite{Singh}. 

Y.I.\ acknowledge A.~Fujimori for valuable comments. 
This work was supported by JST, TRIP.


\begin{thebibliography}{99}
\bibitem{lowT_P05}
	Here, we recorded the spectra of on a fractured surface of P05 
	with an energy resolution of 1 meV. 
	One out of five fracturing resulted in spectra exhibiting 
	superconducting gap and pseudogap features.
\end{thebibliography}
\end{document}